\documentclass[twocolumn]{aastex631}
\usepackage{lmodern}
\newcommand{\uat}[2]{#1 (#2)}

\usepackage{CJK}
\usepackage{amsbsy}
\usepackage{hyperref}

\received{April 15, 2025}
\shorttitle{The optimal Pad\'{e}  polynomial for reconstruction of luminosity distance}
\shortauthors{Yu et al.}
\graphicspath{{./}{figures/}}

\begin{document}
\begin{CJK*}{UTF8}{gbsn}
\title{The optimal Pad\'{e}  polynomial for reconstruction of luminosity distance   based on   10-fold cross-validation}

\author[0000-0002-9849-1762]{Bo Yu(于波)}
\affiliation{School of Mathematics and Big Data, Dezhou University, Dezhou 253023,  China}
\affiliation{Institute for Astronomical Science, Dezhou University, Dezhou 253023, China}
\affiliation{School of Physics and Astronomy, Beijing Normal University, Beijing 100875, China}

\author{Wen-Hu Liu(刘文虎)} 
\affiliation{School of Mathematics and Big Data, Dezhou University, Dezhou 253023,  China}

\author[0000-0002-1265-5747]{Xiaofeng Yang(杨晓峰)}
\affiliation{Xinjiang Astronomical Observatory, Chinese Academy of Sciences, Urumqi 830011, China}
\affiliation{ School of Astronomy and Space Science,
University of Chinese Academy of Sciences, Beijing, 100049, China}
\affiliation{ Key Laboratory of Radio Astronomy and Technology,
Chinese Academy of Sciences, Beijing ,100101, China}
\affiliation{Xinjiang Key Laboratory of Radio Astrophysics, Urumqi, 830011, China}

\author[0000-0002-3363-9965]{Tong-Jie Zhang (张同杰)}
\affiliation{School of Physics and Astronomy, Beijing Normal University, Beijing 100875, China}
\affiliation{Institute for Frontiers in Astronomy and Astrophysics, Beijing Normal University, Beijing 102206, China}
\affiliation{Institute for Astronomical Science, Dezhou University, Dezhou 253023, China}

\author[0000-0003-4207-1694]{Yanke Tang(唐延柯)}
\affiliation{College of Physics and Electronic Information, Dezhou University, Dezhou 253023, People’s Republic of China}

\correspondingauthor{Tong-Jie Zhang}
\email{tjzhang@bnu.edu.cn}



\begin{abstract}

The cosmography known as the Pad\'{e}  polynomials has been widely used in the
 reconstruction of luminosity distance, and the orders of Pad\'{e}  polynomials influence the reconstructed result derived from Pad\'{e}  approximation.
In this paper, we present a more general scheme of selecting optimal Pad\'{e}  polynomial for reconstruction of luminosity distance   based on   10-fold cross-validation. Then  the proposed scheme is applied to Pantheon+ dataset. The  numerical results clearly indicate that the proposed procedure has a remarkable ability to distinguish Pad\'{e} approximations with different orders for the reconstruction of the luminosity distance. We conclude that the (2,1) Pad\'e approximation is  the optimal approach that can well  explain Pantheon+ data at low and high redshifts. Future applications of this scheme could help choose the optimal model that is more suitable for cosmological observation data at hand and gain a deeper understanding of the universe.

\end{abstract}



\section{Introduction} \label{sec:intro}

Recent astronomical observations\citep{1998AJ116.1009R,Perlmutter1998MeasurementsO,2009Baryon} (such as supernovae (SN Ia) and baryon acoustic oscillations (BAO), etc.) indicate that the current universe is undergoing a phase of accelerated expansion. For the purpose of explaining this mysterious phenomenon and the physical principles behind it, many cosmological models have been proposed\citep{Li_2011}. Based on the fact that the cosmology distance depends on the cosmological model (parameters) and redshift, it is very important to accurately and effectively calculate the cosmological distances (e.g., luminosity distance and angle diameter distance, etc.), for testing various cosmological model based on the observation data, restrictions on cosmology parameters, and deep understanding of the formation and evolution of the universe. From the perspective of observation, the luminosity distance is an important cosmological distance. For example, we can use luminosity distance observations of supernovae to constrain the state equation of dark energy. Therefore, it is of great significance for us to reconstruct luminosity distances with reconstruction methods based on observation\citep{10.1093/mnras/staa871,Jesus_2025,YU2020100734,Wei_2014}.

Generally speaking, reconstruction methods of luminosity distance can be divided into two categories: parametric and non-parametric reconstructions\citep{2016MNRAS.460..273M}. In recent years, non-parametric reconstruction has received great attention due to its advantage of assuming only the homogeneity and isotropy of the Universe.  Among the many possible nonparametric reconstruction strategies, Pad\'{e} approximations stabilize the behavior of the cosmographic series at high redshifts, reducing uncertainties of the fitting coefficients and improving the predictive power, which is considered the most promising method\citep{PhysRevD.90.043531,2019MNRAS.484.4484C,10.1093/mnras/staa871,Jesus_2025}.  \citet{PhysRevD.90.043531} actually refined the cosmic bounds by using  the Pad\'{e}  approximation method. \citet{2019Dynamic} applied the Pad\'{e} approximation to approximate the Hubble parameters, which works better than the standard Taylor series approximation for $z>1$. Since rational polynomial offers a better convergence at high redshifts, \citet{PETRECA2024101453} used Pad\'{e} polynomial to construct  $f(z)$CDM.  However, there is an important problem of the choice of orders of Pad\'{e}  polynomials in the reconstruction of the luminosity distance by using Pad\'{e}  polynomials, which may result in  the divergence problem and reduce its predictive ability. On the one hand, when the orders of Pad\'{e}  polynomials  too high and the number of  coefficients is too many, the uncertainties of the free coefficients increases; on the other hand,when the number of  coefficients is too little, the accuracy of the  Pad\'{e} approximation of  luminosity distance  reduces. Based on exact cosmological models, \citet{PhysRevD.90.043531} gave the principles for selecting the orders of  Pad\'{e} approximation of luminosity distance:(1)the Pad\'{e} approximation must not exhibit singularities within any redshift range and must be positive definite; (2)the orders of the numerator and denominator should be close, with the former slightly larger than the latter.\citet{2016EPJC...76..281Z}  considered a moderate order($2,2$) in the work of Pad\'e approximation of the luminosity distance.  \citet{2019MNRAS.484.4484C} obtained  a similar principle for choosing Pad\'{e} order that the order of Pad\'{e} polynomials for numerator and denominator should be close to each other, using the Pad\'{e} polynomial to approximate the Hubble parameters.
 \citet{10.1093/mnras/staa871} pointed out that in order to explain the low and high redshift data, the order($2,1$) of Pad\'{e} approximation is statistically the optimal method  and demonstrated this statement based on the constructions of Padé series and on mathematical rules derived from the degeneracy among coefficients;  and the order($3,2$) of Pad\'{e} approximation cannot be fully excluded at high redshifts. \citet{PETRECA2024101453} noted a better stability in polynomials with the denominator order is lower than the numerator one and considered the stability of Pad\'{e} approximation with order($2,1$), Pad\'{e} approximation with order($2,2$) and Pad\'{e} approximation with order($3,2$).
In \cite{YU2021100772}, Bo YU suggested that the best way to solve this problem of
orders  choice of rational polynomial is to analyze the  Pad\'{e} polynomials with different orders.

As far as we know, there is no specific procedure for the orders choice of approximation for luminosity distance only based on  astronomical observations. Fortunately, cross validation(CV)\citep{1974The,ZHANG201595} provides a general tool to solve the problem of  selecting  the orders of the Pad\'{e} polynomial when reconstructing luminosity distance for cosmological observation data at hand. Basically, based on data partitioning, a portion of the data is used to fit each Pad\'{e} polynomial with different orders, and the remaining data is used to evaluate the predictive performance of the Pad\'{e} polynomial  based on validation errors , and the optimal Pad\'{e} polynomial with the best overall performance is chosen.

In this paper, based on cross validation(CV), we propose a specific procedure for selecting optimal order of Pad\'{e} polynomial by evaluation criterion, such as MSE(Mean Square Error), AIC ( Akaike Information Criterion\citep{1973Information}) and BIC(Bayesian Information Criterion\citep{schwarz1978}), when reconstructing luminosity distance for  1701 updated observations of Pantheon supernovae type Ia (SNeIa). The structure  of this paper is organized as follows: Sec. \ref{sec:data} introduces the measurement of  1701 updated observations of Pantheon supernovae type Ia (SNeIa);
In Sec.\ref{sec:opdl}, we propose a specific procedure for the orders choice of Pad\'{e} approximation for luminosity distance based on  cross validation;
Sec.\ref{sec:Results} present a comparison of orders choice for Pad\'{e} approximation of SNe Ia by three evaluation criteria (MSE, AIC and BIC; finally, we discuss the results in Sec.\ref{sec:Results},  and
the conclusions of this study is provided in Sec.\ref{sec:conclusion}. 

\section{ DATA} \label{sec:data}

The Pantheon+ dataset\citep{Scolnic2022} is a comprehensive collection of cosmological observations from Type Ia supernovae (SNe Ia), which is considered as critical "standard candles" for measuring cosmic distances. By integrating an extended dataset of supernovae (SNe) that includes those associated with galaxies for which Cepheid distances have been measured, the Pantheon+ dataset offers significant enhancements and improvements, consisting of 1701 light curves of 1550 unique covering a redshift range from 0.001 to 2.2613 in terms of the distance modulus $\mu_{obs}{(z_i)}$. Compared with the original Pantheon compilation\citep{Scolnic_2018}, the most notable change of Pantheon+ dataset is significant increase in sample size, especially the significant increased number of SNe with redshifts below 0.01. By \citet {2024Exploring}, the  $\chi^2$ from ''Pantheon+ dataset" 1071 SNe Ia data can be defined as
\begin{equation}
\chi^2_{SNe} = \Delta \mu^T (C_{\mathrm{Sys}+\mathrm{Stat}}^{-1}) \Delta \mu
\end{equation}
where $C_{\mathrm{Sys}+\mathrm{Stat}}$ is the covariance matrix of the Pantheon+ dataset, which incorporates the combination of systematic and statistical uncertainties; and $\Delta\mu$ indicates the observed distance residual that is defined as $\Delta \mu_k = \mu_k - \mu_{\text{th}}(z_k)$, where $\mu_k$ represents the distance modulus of the $k^{th}$ SNe, and the theoretical distance modulus $\mu_{\text{th}}$ is defined as:
\begin{equation}
\mu_{\text{th}}=5\log_{10}{(\frac{{d}_L(z)}{Mpc})}+ 25
\end{equation}
where  $d_L$  indicates the  luminosity distance of non-parametric reconstruction  in megaparsecs $Mpc$.

\section{Optimal Pad\'{e} approximation for luminosity distance based on  cross validation} \label{sec:opdl}

\subsection{Pad\'{e}  polynomial}
\label{sec:3.1}

As mentioned in \citet{Wuytack1979Pade},  Pad\'e approximation is generally considered a good choice for   approximating of a given function $f(z)$ by a rational polynomial with an order($m$,$n$)
\begin{equation}
P_{m,n}(z)=\frac{a_0+a_1z+\dots+a_mz^m}{1+b_1z+\dots+b_nz^n}
\end{equation}
To ensure the stability of  cosmographic expansions at high redshifts, it is possible as one assumes rational approximations  with an order($m$,$n$) of a given cosmological observable\citep{10.1093/mnras/staa871}. Generally, the Pad\'e approximation is built up from the standard Taylor expansion and is used to lower divergences at high redshifts.

Let
 us suppose that $f(z)$ is expanded in Taylor series:
 \begin{equation}
f(z)=\sum\limits_{k=0}^{m+n+1}{d_kz^k}
\end{equation}
Then the coefficients of Pad\'e approximation with order($m$,$n$) can be determined by using the following formulas:
\begin{eqnarray}
&P_{m,n}(z)=f(0) \nonumber\\
&P_{m,n}^{'}(z)=f^{'}(0) \nonumber\\
&\vdots \nonumber \\
&P_{m,n}^{(m+n+1)}(z)=f^{(m+n+1)}(0)
\label{con:Padeecc}
\end{eqnarray}

According to \cite{10.1093/mnras/staa871}, Pad\'e approximations of luminosity distance with order($2$,$1$)
$P_{2,1}$, order($2$,$2$) $P_{2,2}$ , order($3$,$1$)
$P_{3,1}$ and order($3$,$2$)
$P_{3,2}$ can be reconstructed as follows:

\begin{equation}
P_{2,1}(z) = \frac{1}{H_0} \left[ \frac{z\left(6(-1+q_0) + (-5-2j_0 + q_0(8+3q_0))z\right)}{-2(3+z+j_0 z) + 2q_0(3+z+3q_0 z)} \right]
\end{equation}
\begin{eqnarray}   
&P_{2,2}(z) = \frac{1}{H_0} (6z (10 + 9z - 6q_0^3 z+ s_0 z - 2q_0^2(3+7z)\nonumber\\ & - q_0(16+ 19z)  + j_0(4 + (9+6q_0)z)))/(60 + 24z +\nonumber\\& 6s_0 z - 2z^2 + 4j_0^2 z^2- 9q_0^4 z^2 - 3s_0 z^2  + 6q_0^3 z(-9 + \nonumber\\&4z) + q_0^2(-36 - 114z + 19z^2)+ j_0(24 + 6(7+8q_0)z \nonumber\\&+ (-7-23q_0 + 6q_0^2)z^2)+ q_0(-96 - 36z + \nonumber\\&(4+3s_0)z^2))
\end{eqnarray}
\begin{eqnarray}
    &P_{3,1}(z) = \frac{1}{H_0} (z(z^3(10j0q0 + 5j0 - 15q0^3\nonumber\\& - 15q0^2 - 2q0 + s0 + 2) + 3z^2(j0 -\nonumber\\& 3q0^2 - q0 + 1) - 6z(q0 - 1) +\nonumber \\&6))/(3(z(-j0 + 8q0 + 1) + 2)
\end{eqnarray} 

\begin{eqnarray}
&P_{3,2}(z) =  \frac{1}{H_0} (z(-120 - 180s_0 - 156z - 36l_0z \nonumber\\&- 426s_0z - 40z^2 + 80j_0^3z^2 - 30l_0z^2 - 135q_0^6z^2 -\nonumber\\& 210s_0z^2 + 15s_0^2z^2 -270q_0^5z(3 + 4z)+ 9q_0^4(-60
 \nonumber\\&+ 50z+ 63z^2) + 2q_0^3(720 + 1767z + 887z^2) +\nonumber\\& 3j_0^2(80 + 20(13 + 2q_0)z 
 + (17 + 40q_0 - 60q_0^2)z^2) + \nonumber\\&6q_0^2(190 + 5(67 + 9s_0)z + (125 + 3l_0 + 58s_0)z^2)- \nonumber\\&6q_0(s_0(-30 + 4z + 17z^2) - 2(20 + (31 + 3l_0)z + \nonumber\\&(9 + 4l_0)z^2)) + 6j_0(-70 + (-127 + 10s_0)z + \nonumber\\& 45q_0^4z^2+  (-47 - 2l_0 + 13s_0)z^2  + 5q_0^3z(30 + 41z)\nonumber\\& - 3q_0^2(-20 + 75z + 69z^2) + 2q_0(-115 - 274z  +\nonumber\\& (-136 + 5s_0)z^2))))/(3(-40 - 60s_0 - 32z - \nonumber\\&12l_0z - 112s_0z - 4z^2   + 40j_0^3z^2 - 4l_0z^2 \nonumber\\&- 135q_0^6z^2 - 24s_0z^2+ 5s_0^2z^2 - 30q_0^5z(12 + 5z) +\nonumber\\& 3q_0^4(-60 + 160z+ 71z^2)+ j_0^2(80 + 20(11 + 4q_0)z \nonumber\\&+ (57 + 20q_0 - 40q_0^2)z^2) + 6q_0^3(80 + 188z + \nonumber\\&(44 + 5s_0)z^2) + 2q_0^2(190 + 20(13 
       + 3s_0)z + (46 + 6l_0 +\nonumber\\& 21s_0)z^2) + 4q_0(20 
       + (16 + 3l_0)z + (2 + l_0)z^2\nonumber\\& + s_0(15 - 17z - 9z^2)) + 2j_0(-70 + 2(-46 + 5s_0)z 
       + 90q_0^4z^2 \nonumber\\&+ (-16 - 2l_0 +3s_0)z^2 +15q_0^3z(12 
       + 5z) + q_0^2(60 - 370z- \nonumber\\&141z^2)
       + 2q_0(-115 - 234z + 2(-26 + 5s_0)z^2))))
\end{eqnarray} 
where,
\begin{eqnarray}
& q_0 = \frac{3}{2}\Omega_{m0} - 1, \nonumber\\
  & j_0 = 1,\nonumber \\
  & s_0 = 1 - \frac{9}{2}\Omega_{m0},\nonumber \\
  & l_0 = 1 + 3\Omega_{m0} + \frac{27}{2}\Omega_{m0}^2, \nonumber\\
  & m_0 = -\frac{81}{4}\Omega_{m0}^3 - 81\Omega_{m0}^2 - \frac{27}{2}\Omega_{m0} + 1\nonumber    
\end{eqnarray}

Naturally, it is important to confront the mentioned above Pad\'e polynomials  of different orders  with the observational data, and see which one is most suitable. Since Pantheon+  data are directly related to the luminosity distance, we can apply them to constrain  Pad\'e polynomials  with different orders .

\subsection{Cross validation and selection criteria }
\label{sec:3.2}
As mentioned in \citet{ZHANG201595}, cross validation (CV)provides a general tool to select the best candidate model for data at hand. In this paper we apply CV to solve the problem of selecting  the orders of the Pad\'{e} polynomial, when reconstructing luminosity distance using  Pantheon+  dataset. Based on Pantheon+  data partitioning, a portion of the data is used to fit each Pad\'{e} polynomial with different orders, and the remaining data is used to evaluate the predictive performance of each Pad\'{e} polynomial  by validation errors , and the Pad\'{e} polynomial with the best overall performance is selected. In the literature, there are many recommendations on the data splitting methods for CV  and $k$-fold CV seems to be a favorite of
many researchers\citep{10.1214/09-SS054}.

In $ k$ -fold CV, the Pantheon+ data are randomly divided  into $k$ equal-size subsets. In turn each of the $k$ subsets is retained as the validation set, while the rest \(k - 1\)  folds use as the training set, and the average prediction error of each candidate Pad\'{e} polynomial is obtained. In the absence of prior information about the real model of data, there is an important issue on inappropriate use of model selection criteria may lead to poor results. 

Generally speaking, model selection aims to select the best statistical  model from a set of candidates, balancing its goodness of fitting and complexity, and key criteria usually include: Mean Square Error (MSE), Akaike information criterion (AIC) and Bayesian Information Criterion (BIC).

In statistics, the Mean Square Error (MSE) is a widely used metric to evaluate the accuracy of a predictive model, which quantifies the average squared difference between predicted values ($\hat y_i$) and real observed values ($y_i$)\citep{1922On}. MSE is given as: 

\begin{equation}
    \text{MSE} = \frac{1}{n} \sum_{i=1}^{n} (\hat{y}_i - y_i)^2
\end{equation}
where $n$ is sample size.

The Akaike information criterion (AIC) \citep{1973Information} is defined by

\begin{equation}
AIC= 2m-2\ln (L_{max})
\end{equation}
where $L_{max}$ is the maximum likelihood function, and $m$ is the number of free coefficients  for  candidate model.

Bayesian Information Criterion (BIC)\citep{schwarz1978} is given as
\begin{equation}
BIC= m\ln{n}-2\ln (L_{max})
\end{equation}
where  $m$ is the number of free coefficients  for  candidate model, $n$ is sample size, and $L_{max}$ is the maximum likelihood function, respectively.

\subsection{Optimal Pad\'{e} approximation scheme for luminosity distance }
\label{sec:3.3}

In order to avoid the impact of different $k$-fold cross validation on the results, we provide the  Pad\'e polynomials  with different orders based on different  $k$-fold for  Pantheon+ data separately.

The implementation of optimal Pad\'{e} approximation scheme for luminosity distance can be summarized as follows:

Step 1: Based on different $k$-fold CV ( such as $k=2, 5, 10, 20$, and $50$), the Pantheon+ data are randomly divided  into $k$ equal-size subsets. In turn each of the $k$ subsets is retained as the validation set, while the rest \(k - 1\)  folds use as the training set, and the average of MSE of each  Pad\'{e} polynomial with  different orders is obtained.

Step 2:  We can initially select candidate Pad\'{e} polynomials by the average of MSE  in different $k$-fold. The fitting results for distance modulus of candidate Pad\'{e} polynomials are shown in Figure\ref{fig:pade_diff}.

Step 3: Repeat the operation of step 1, the averages of MSE, AIC, and BIC for each candidate Pad\'{e} polynomial are obtained, and these results are shown in Figure\ref{fig:Mean_MSE_K_folds}, Figure\ref{fig:Mean_BIC_K_folds} and Figure\ref{fig:Mean_AIC_K_folds}, respectively.

Step 4: According to the model selection criteria obtained in step 3, the most suitable Pad\'{e} polynomial for the Pantheon+ data is selected.

Step 5: By minimizing the $\chi^2$ from ''Pantheon+ dataset''1071 SNe Ia data, we get the optimal Pad\'{e} approximation of luminosity distance.

\section{Results }
\label{sec:Results}

As mentioned in Step 1, based on different $k$-fold CV ( such as $k=2, 5, 10, 20, 50$), the average of MSE of each  Pad\'{e} polynomial with  different orders is obtained. According to the criterion of the average of MSE less than 0.05, we  can preliminarily select candidate Pad\'{e} polynomials with order($2$,$1$)
$P_{2,1}$, order($2$,$2$) $P_{2,2}$, order($3$,$1$)
$P_{3,1}$ and order($3$,$2$)
$P_{3,2}$ in 2,5,10,20 and 50 fold CV, respectively. The fitting results for distance modulus of candidate Pad\'{e} polynomials are shown in Figure\ref{fig:pade_diff}. From Figure\ref{fig:pade_diff}, the preliminary candidate Pad\'{e} polynomials fit Pantheon+ data well,  $P_{2,1}$ and $P_{3,2}$ work better than that of $P_{3,1}$ and $P_{2,2}$.

\begin{figure*}[ht!]
\includegraphics{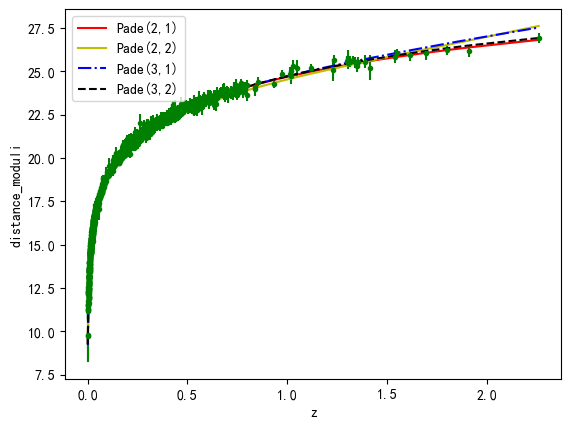}
\caption{Hubble diagram for Pantheon+ Sample. The fitting results for distance modulus of  preliminary candidate Pad\'{e} polynomials \label{fig:padediff}}
\end{figure*}

\begin{figure*}[ht!]
\includegraphics{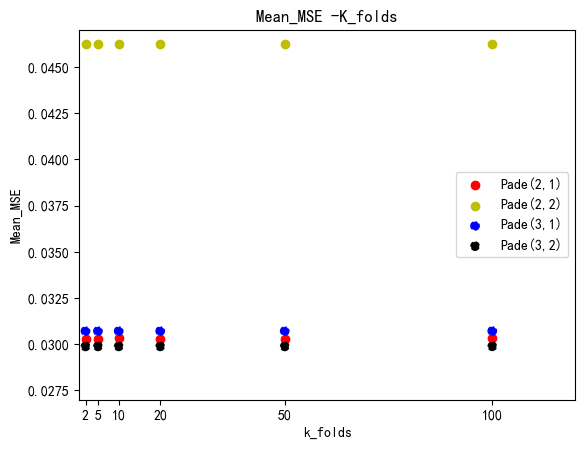}
\caption{Hubble diagram for Pantheon+ Sample. The fitting results for distance modulus of  preliminary candidate Pad\'{e} polynomials \label{fig:MeanMSEKfolds}}
\end{figure*}

\begin{figure*}[ht!]
\includegraphics{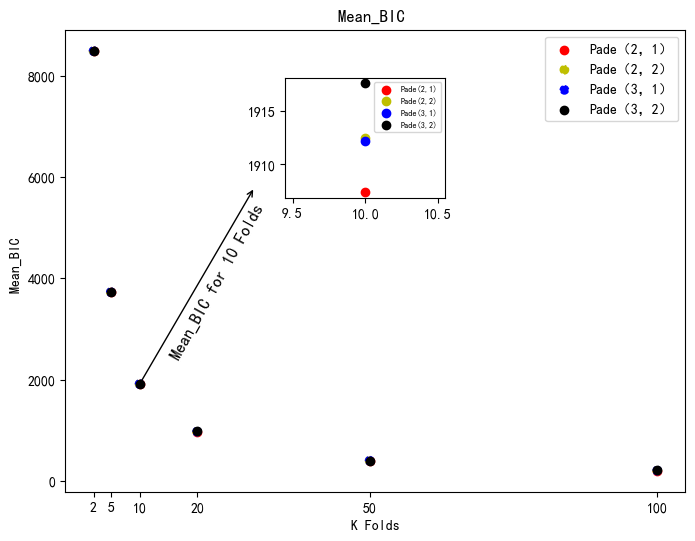}
\caption{The average of BIC of preliminary candidate Pad\'{e} polynomials  in 2,5,10,20 and 50 fold CV, respectively. The difference of average of BIC  in 10-fold CV
that is amplified  is shown in inset. \label{fig:MeanBICKfolds}}
\end{figure*}

\begin{figure*}[ht!]
\includegraphics{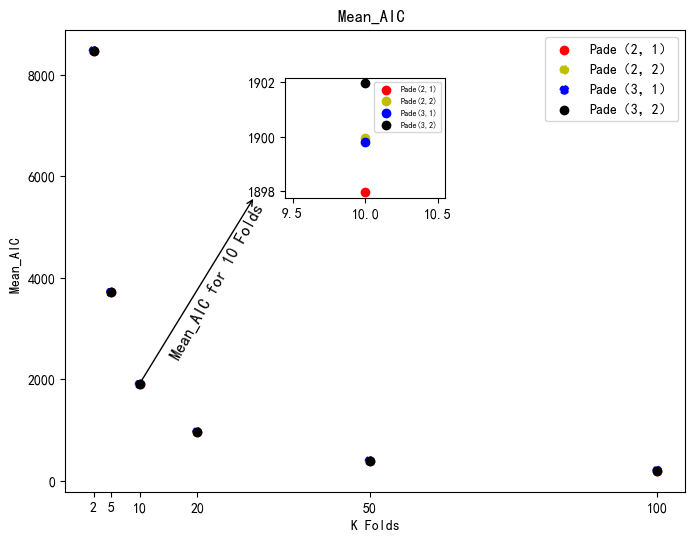}
\caption{The average of AIC of preliminary candidate Pad\'{e} polynomials  in 2,5,10,20 and 50 fold CV, respectively. The difference of average of  AIC in 10-fold CV
that is amplified  is shown in inset.\label{fig:MeanAICKfolds}}
\end{figure*}

Then  according to Step 3, we obtain  the averages of MSE, AIC, and BIC for each candidate Pad\'{e} polynomial, see Figure\ref{fig:MeanMSEKfolds}, Figure\ref{fig:MeanBICKfolds} and Figure\ref{fig:MeanAICKfolds}, respectively.
In Figure\ref{fig:MeanMSEKfolds}, we can get that  $P_{3,2}$ and $P_{2,1}$ are in good agreement with cosmological observation data  from  the perspective of average of MSE.  Figure\ref{fig:MeanBICKfolds} and Figure\ref{fig:MeanAICKfolds} indicate that the difference of average of BIC and AIC is not significant. In practical applications, the model selection using 10-fold CV has been widely accepted by researchers \citep{1995A}. By the difference of average of BIC and AIC in 10-fold CV that is amplified in  insets of Figure\ref{fig:MeanBICKfolds} and Figure\ref{fig:MeanAICKfolds}, we obtain that Pad\'{e} polynomial with order ($2$,$1$) fits the data best.

\begin{figure*}[ht!]
\includegraphics{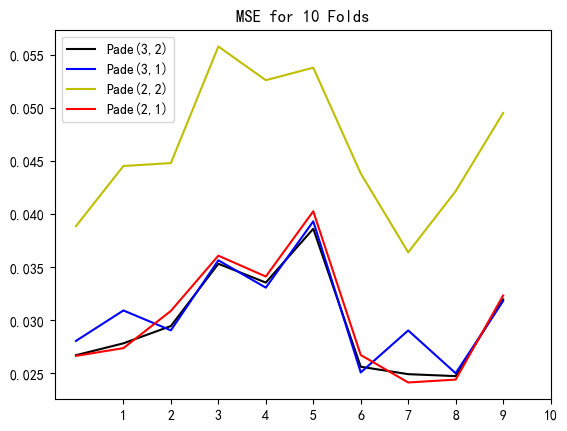}
\caption{The MSE of preliminary candidate Pad\'{e} polynomials  in 10-fold CV.\label{fig:MSE10folds}}
\end{figure*}

\begin{figure*}[ht!]
\includegraphics{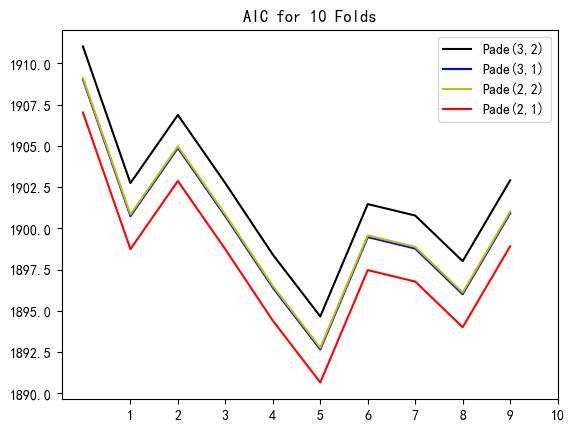}
\caption{The AIC of preliminary candidate Pad\'{e} polynomials  in 10-fold CV.\label{fig:AIC10folds}}
\end{figure*}

\begin{figure*}[ht!]
\includegraphics{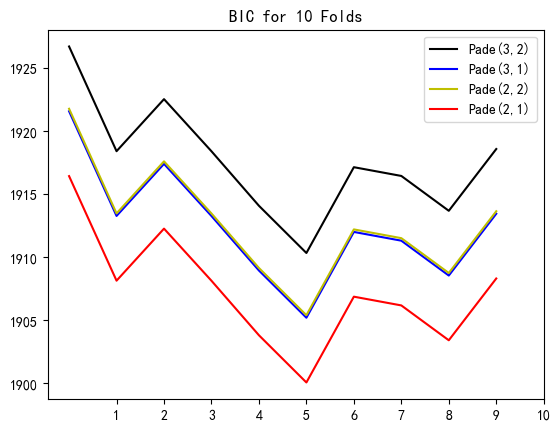}
\caption{The BIC of preliminary candidate Pad\'{e} polynomials  in 10-fold CV.\label{fig:BIC10folds}}
\end{figure*}

The MSE , AIC and BIC of candidate Pad\'{e} polynomials in 10-fold CV have been used for the more detailed and direct display of the difference among them in order for the us to compare and chose the optimal Pad\'{e} polynomial more conveniently, see Figure\ref{fig:MSE10folds}, Figure\ref{fig:AIC10folds} and Figure\ref{fig:BIC10folds}, respectively. Overall, it is clear that  Pad\'{e} polynomial with order ($2$,$1$) fits the data best by MSE , AIC and BIC based on 10-fold CV.

For the purpose of getting the best fits of  Pad\'{e} polynomial with order ($2$,$1$) , we apply Markov Chain Monte Carlo (MCMC) code emcee\citep{Foreman_Mackey_2013}  on ''Pantheon+ ''dataset likelihood. Table \ref{tab:p21contour} reports the numerical results for parameters of  Pad\'e approximations of luminosity distance with order(2,1). The marginal distribution function in one-dimensional parameter space, and 1$\delta$, and 2$\delta$  contour in the two-dimensional parameter space for the best fitting from  "Pantheon+" dataset are shown in Figure \ref{fig:p21emcee}.

\begin{table}[htbp]
    \centering
    \caption{MCMC results at the 68.3\%  C.L for Pad\'e approximations of luminosity distance with order(2,1) from ''Pantheon+ ''dataset. }
    \label{tab:p21contour}
    \begin{tabular}{cccc}
        \hline
        Model & $q_0$ & $j_0$& $ H_{0}$ \\
        \hline
        $P(2,1)$ & $-0.52_{-0.12}^{+0.15}$ &$1.5_{-1.2}^{+1.1}$ &  $71.84_{-0.19}^{+0.25}$\\
        \hline
    \end{tabular}
\end{table}

\section{Conclusions and discussion}
\label{sec:conclusion}
Among the many possible nonparametric reconstruction strategies based on given cosmological observation data, Pad\'{e} approximation is considered the most promising method due to its advantages of  stabilizing the behavior of the cosmographic series at high redshifts, reducing uncertainties of the fitting coefficients and improving the predictive power.  It is worth noting that the order selection  of Pad\'{e}  polynomials in the reconstruction of the luminosity distance is very important, which may reduce its predictive ability and result in  the divergence problem. In addition, the approach of using 10-fold cross validation for model selection has been widely accepted.

\begin{figure*}[h!]
\includegraphics{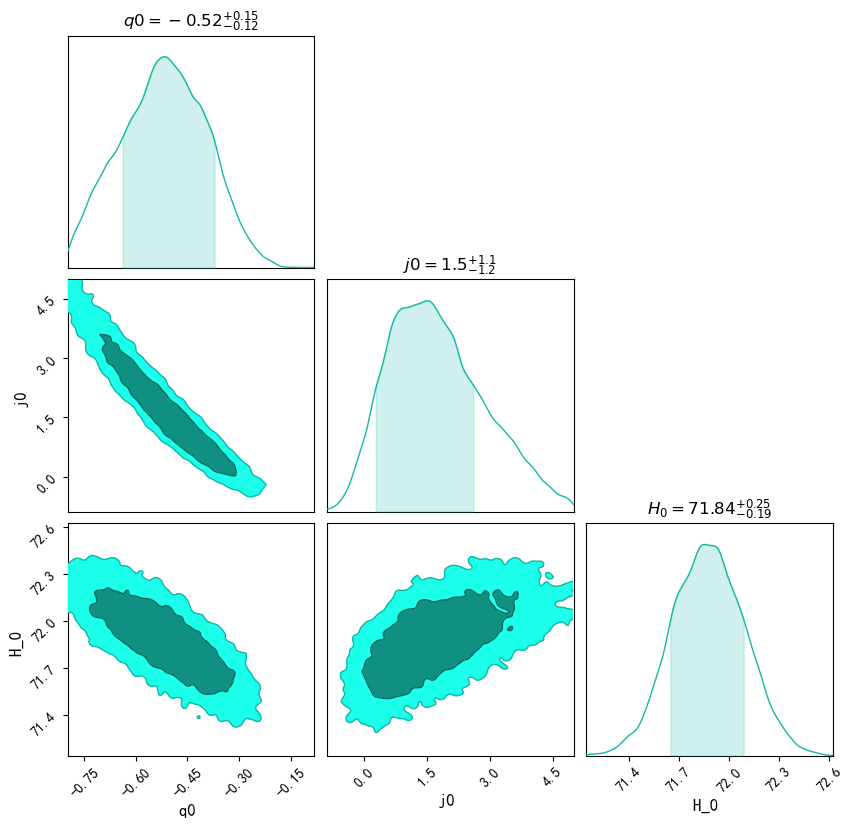}
	\caption{ The  contour for 1$\delta$, and 2$\delta$  in the two-dimensional parameter spaces of P(2,1), and  the marginal distribution function in one-dimensional parameter space.\label{fig:p21emcee}}
\end{figure*}

In this paper, based on 10-fold CV, a specific scheme  for selecting optimal Pad\'{e} approximation of luminosity distance for given cosmological observation data is proposed. To reduce the impact of randomness in the  data partitioning, we use the average of MSE of each  Pad\'{e} polynomial in k-fold CV, see Figure\ref{fig:MeanMSEKfolds}. For the elucidative purposes, we make the comparison of Pad\'e approximations with different orders by the AIC and BIC in 10-fold CV  in Figure\ref{fig:AIC10folds} and Figure\ref{fig:BIC10folds}. The  numerical results clearly indicate that the proposed procedure has a remarkable ability to distinguish Pad\'{e} approximations with different orders for the reconstruction of the luminosity distance. Combined with Table\ref{tab:p21contour} and Figure \ref{fig:p21emcee}, it is shown that optimal Pad\'{e} polynomial with order ($2$,$1$) can well  explain ''Pantheon+ '' data at low and high redshifts, which is consistent with the S. Capozziello 's conclusion in \citet{10.1093/mnras/staa871}.

In summary, this paper proposes a scheme for selecting the order of Pad\'{e} approximation only based on cosmological observation datasets. To verify the scientific validity of the proposed scheme, it is essential to compare the results obtained by our scheme with the ones previously acquired using Pad\'{e} expansions. In \citep{PhysRevD.90.043531},  the authors obtained the statement that  Pad\'{e} approximates whose numerator and denominator have similar degrees seem to be preferred，based on the Pad\'{e} approximates for exact cosmological models and corresponding fitting results， and pointed out that the order($2,1$) of Pad\'{e} approximation, order($3,1$) of Pad\'{e} approximation  order($2,2$) of Pad\'{e} approximation are the ones that draw the best samples, with the narrowest dispersion.  By using the Pad\'{e} polynomial to approximate the Hubble parameters, \citet{2019Dynamic} obtained a principle for choosing Pad\'{e} order: under the condition of ensuring convergence, the Pad\'{e} polynomials should evolve smoothly in all redshift ranges used for cosmological analysis, and ensure that the Hubble parameters are positive; in addition, the order of Pad\'{e} polynomials for numerator and denominator should be close to each other, which is similar to \citet{PhysRevD.90.043531}. From the perspective of the constructions of Pad\'{e} series and the mathematical rules derived from the degeneracy among coefficients, \citet{10.1093/mnras/staa871} singled out  the most suitable rational approximation is the one that minimizes the number of parameters of denominator, and
 Pad\'{e} approximation with order($2,1$) is statistically the optimal method to explain low- and high-redshift data. In  terms of stability and goodness of fitting, \citet{PETRECA2024101453} considered that Pad\'{e} polynomials with the denominator order is lower than the numerator one are preferred. In summary, based on the results of previous using Pad\'{e} polynomials, we can achieve the following principle for selecting the order of the Pad\'{e} polynomial: (1) the order of Pad\'{e} polynomials for numerator and denominator should be close to each other,and the ones with the denominator order is lower than the numerator one are preferred; (2) 
the Pad\'{e} polynomials should evolve smoothly in all redshift range , and should be positive.  Based on combined with Table\ref{tab:p21contour} and Figure \ref{fig:p21emcee}, we conclude that optimal Pad\'{e} polynomial with order ($2$,$1$) can well  explain ''Pantheon+ '' data at low and high redshifts, which  is  consistent with the two principles  selecting the order of the Pad\'{e} polynomial mentioned above. In other words, we provide another approach that relies solely on the cosmological observations  dataset to select the order of Pad\'{e} polynomials, and can obtain results which are consistent with previous ones using Pad\'{e} approximation.

In addition, the scheme  presented in this paper can be applied to other cosmological observational datasets, such as Cosmic Chronometers (CC) dataset. However, the conclusion of this paper has some limitations: (1) The optimal  Pad\'{e} polynomial chosen in this study  depend only on the  ''Pantheon+'' data. (2) The choice of data splitting ratio may affect the selection of the order of the optimal Pad\'e polynomial. Future works will focus on how to generalize the afore mentioned scheme by also using other cosmological observational datasets, and make new progress in our understanding of the universe. 


\begin{acknowledgments}
We thank Anonymous Referees for their valuable comments
for revising and improving earlier draft of our manuscript. This work was supported by National Key R\&D Program of China，No.2024YFA1611804,  the China Manned Space Program with grant No. CMS-CSST-2025-A01, and National SKA Program of China,
 No.2022SKA0110202, 2022SKA0110203.

\end{acknowledgments}


\bibliography{ref}{}
\bibliographystyle{aasjournal}


\end{CJK*}
\end{document}